\documentclass[12pt]{iopart}

\usepackage{iopams}
\usepackage{setstack}
\usepackage{graphicx}
\usepackage{bm}
\usepackage{hyperref}
\usepackage{verbatim}
\hypersetup{colorlinks=false}

\begin{document}

\title[Analytic solutions in the dyon black hole with a cosmic string]{Analytic solutions in the dyon black hole with a cosmic string: scalar fields, Hawking radiation and energy flux}

\author{H. S. Vieira$^{1,2}$, V. B. Bezerra$^{1}$ and G. V. Silva$^{1}$}

\address{$^{1}$ Departamento de F\'{i}sica, Universidade Federal da Para\'{i}ba, Caixa Postal 5008, CEP 58051-970, Jo\~{a}o Pessoa, PB, Brazil}
\address{$^{2}$ Centro de Ci\^{e}ncias, Tecnologia e Sa\'{u}de, Universidade Estadual da Para\'{i}ba, CEP 58233-000, Araruna, PB, Brazil}

\ead{horacio.santana.vieira@hotmail.com, valdir@fisica.ufpb.br and gislainevs@hotmail.com}

\begin{abstract}
Charged massive scalar fields are considered in the gravitational and electromagnetic field produced by a dyonic black hole with a cosmic string along its axis of symmetry. Exact solutions of both angular and radial parts of the covariant Klein-Gordon equation in this background are obtained, and are given in terms of the confluent Heun functions. The role of the presence of the cosmic string in these solutions is showed up. From the radial solution, we obtain the exact wave solutions near the exterior horizon of the black hole, and discuss the Hawking radiation spectrum and the energy flux.
\end{abstract}





\maketitle


%
%
\section{Introduction}
The knowledge of the behavior of different fields which interact with the gravitational field of black holes, can give us, in principle, some relevant informations about the physics of these objects. In particular, the scalar field constitutes one of these fields whose behavior will be studied in this paper taking into account a background spacetime generated by a rotating black hole with electric and magnetic charges through which passes a cosmic string. Along this line of research, the separability of the Klein-Gordon equation has been studied in different black hole backgrounds \cite{CommMathPhys.10.280,PhysRevD.5.1913,ClassQuantumGrav.22.339}. Other studies concerning the behavior of scalar fields in different black holes background as well as their consequences can be found in \cite{PhysRevD.12.2963,PhysRevD.36.1269,PhysRevD.76.084001}.

The topological defect called cosmic string is predicted in some gauge field theories \cite{JPhysAMathGen.9.1387} as a result of phase transitions. It can either form closed loops or extend to infinity, and is characterized by its tension, $G\mu$, where $G$ is the Newton's gravitational constant and $\mu$ is the mass per unit length of the string. The spacetime geometry associated with a straight infinite and infinitely thin cosmic string has a conical structure which means that it is locally flat but not globally, due to the fact that on the localization of the string the curvature tensor has a delta shaped form. The local flatness the spacetime surrounding a cosmic string means that there is no local gravitational force. The section perpendicular to the cosmic string has an azimuthal deficit angle given by $\triangle \phi=8 \pi G \mu$ \cite{PhysRevD.23.852}. However, there exist some interesting gravitational effects associated with the global features of the spacelike section around the cosmic string. Among these effects, a cosmic string can induce a finite electrostatic self-force on an electric charged particle, it can induce the emission of radiation by a freely moving particle, act as a gravitational lens among others \cite{Vilenkin:1994}.

A cosmic string cannot appear, necessarily, as a single object in empty space. In fact, it can appear as part of a larger gravitational system, as for example, a black hole. In which case the cosmic string is included \cite{bezerra} by removing a wedge, that is, by requiring that the azimuthal angle around the axis of symmetry runs over the range $0 < \phi < 2 \pi b$, with $b=1-4\mu$. Then, gluing together the resulting edges we get the spacetime corresponding to a black hole with a cosmic string passing through it. As an example of such generalizations we can mention the Schwarzschild spacetime with a cosmic string \cite{ClassQuantumGrav.13.2663} and the Kerr spacetime with a cosmic string \cite{ClassQuantumGrav.6.1313}.

An interesting phenomenon which corresponds to a spontaneous emission of black body radiation by black holes is the Hawking radiation, which was predicted from the study of the thermal radiation emitted by a spherically symmetric black hole \cite{CommunMathPhys.43.199}. The studies concerning this phenomenon was carried on by using different methods \cite{PhysRevD.11.1404,PhysRevD.13.2188,PhysRevD.15.2738}. In particular, the emission of scalar particles by black holes has been an object of intense investigations in recent years \cite{AnnPhys.350.14,EPL.109.60006}.

Zhao and Zhang \cite{KexueTongbao.29.1303} by introducing the tortoise coordinate to discuss the Hawking thermal spectrum from Kerr-Newman-Kasuya black hole. Other approaches include the investigation of the general radiation spectrum by using the tunneling mechanism process adopted by Yang \cite{ChinPhysLett.22.2492} and the one which consider the Hawking radiation using the viewpoint of quantum anomalies considered by He et al. \cite{ChinPhysB.17.2321}. In the methods used by these authors it is not necessary the full solution of the scalar equation.

Our contribution is obtaining the exact solutions of the Klein-Gordon equation for a charged massive scalar field in the Kerr-Newman-Kasuya spacetime (dyonic black hole) with a cosmic string passing through it. These solutions are given in terms of the confluent Heun functions \cite{Ronveaux:1995} and are valid in the the region between the exterior event horizon and infinity, which corresponds to the whole space. If we cancel the magnetic charge, namely, take $Q_{m}=0$, we particularize the solutions to the Kerr-Newman spacetime with a cosmic string which are valid in the whole space, which means from the event horizon to infinity, generalizing a result obtained in the literature \cite{ClassQuantumGrav.23.7063}. Using the obtained solutions for the radial part of the Klein-Gordon equation we study the Hawking radiation. The analytical calculation of the outgoing decay rate or emission rate for charged massive scalar particles in the background under consideration is presented. The resulting spectra are integrated to give the total power emitted or energy flux in the various modes.

This paper is organized as follows. In Section 2, we introduce the metric that corresponds to the Kerr-Newman-Kasuya spacetime with a cosmic string passing through it (dyonic black hole with a cosmic string) using a geometric procedure to introduce this topological defect, and we obtain the thermodynamic relevant parameters to study the Hawking radiation. In Section 3, we write down the covariant Klein-Gordon equation for a charged massive scalar field in the background under consideration, and separate the angular and radial parts. In Section 4, we present the exact solutions of both angular and radial equations. In Section 5, we obtain the solutions for ingoing and outgoing waves, near to the exterior horizon of the dyonic black hole with a cosmic string, and we extend the wave solutions from the outside to the inside of this spacetime. In Section 6, under the condition that the spacetime total energy, total charge, and total angular momentum are conserved, we derive the radiation spectra and the energy flux. Finally, in Section 7 we present our conclusions.
%
%
\section{Kerr-Newman-Kasuya spacetime with a cosmic string}
The metric generated by a black hole with angular momentum per mass, $a=J/M$, electric charge, $Q_{e}$, magnetic charge, $Q_{m}$, and mass (energy), $M$, corresponds to the Kerr-Newman-Kasuya metric \cite{PhysRevD.25.995}, whose line element, in the Boyer-Lindquist coordinates, is given by
\begin{eqnarray}
ds^{2} & = & -\frac{\Delta}{\rho^{2}}(dt-a\sin^{2}\theta\ d\phi)^{2}+\frac{\rho^2}{\Delta}\ dr^{2}+\rho^{2}\ d\theta^{2}\nonumber\\
 & + & \frac{\sin^2\theta}{\rho^2}[(r^2+a^2)d\phi-a\ dt]^{2}\ ,
\label{eq:metrica_Kerr-Newman-Kasuya}
\end{eqnarray}
where
\begin{equation}
\Delta=r^{2}-2Mr+a^{2}+Q_{e}^{2}+Q_{m}^{2}\ ,
\label{eq:Delta_metrica_Kerr-Newman-Kasuya}
\end{equation}
and
\begin{equation}
\rho^{2}=r^{2}+a^{2}\cos^{2}\theta\ .
\label{eq:rho_metrica_Kerr-Newman-Kasuya}
\end{equation}

Now, let us introduce a cosmic string in the Kerr-Newman-Kasuya spacetime. This can be made by a geometric procedure by simply redefining the azimuthal angle $\phi$ in such a way that $\phi \rightarrow b\phi$, and gluing the resulting edges. Doing this, we obtain the metric corresponding to a Kerr-Newman-Kasuya spacetime with a cosmic string passing through it, which can be rewritten as
\begin{eqnarray}
ds^{2} & = & g_{\sigma\tau}dx^{\sigma}dx^{\tau}\nonumber\\
& = & -\frac{\Delta-a^{2}\sin^{2}\theta}{\rho^{2}}\ dt^{2}+\frac{\rho^{2}}{\Delta}\ dr^{2}+\rho^{2}\ d\theta^{2}\nonumber\\
& + & \frac{(r^{2}+a^{2})^{2}-\Delta a^{2}\sin^{2}\theta}{\rho^{2}}b^{2}\sin^{2}\theta\ d\phi^{2}\nonumber\\
& - & 2\frac{(r^{2}+a^{2})-\Delta}{\rho^{2}}a b\sin^{2}\theta\ dt\ d\phi\ ,
\label{eq:metrica_Kerr-Newman-Kasuya_string}
\end{eqnarray}
where the parameter $b$, which codifies the presence of the cosmic string, assumes values in the interval $0 < b < 1$.

From Eq.~(\ref{eq:metrica_Kerr-Newman-Kasuya_string}), we have that the horizon surface equation is obtained from the condition
\begin{equation}
\Delta=(r-r_{+})(r-r_{-})=0\ ,
\label{eq:superficie_hor_Kerr-Newman-Kasuya}
\end{equation}
whose solutions are
\begin{equation}
r_{+}=M+[M^{2}-(a^{2}+Q_{e}^{2}+Q_{m}^{2})]^{1/2}\ ,
\label{eq:sol_padrao_Kerr-Newman-Kasuya_1}
\end{equation}
\begin{equation}
r_{-}=M-[M^{2}-(a^{2}+Q_{e}^{2}+Q_{m}^{2})]^{1/2}\ ,
\label{eq:sol_padrao_Kerr-Newman-Kasuya_2}
\end{equation}
and correspond to the event and Cauchy horizons.

The surface gravity of the black hole or the gravitational acceleration on the background horizon surface, $r_{+}$, is given by
\begin{equation}
\kappa_{+} \equiv \frac{1}{2}\frac{1}{r_{+}^{2}+a^{2}}\left.\frac{d\Delta}{dr}\right|_{r=r_{+}}=\frac{1}{2}\frac{r_{+}-r_{-}}{r_{+}^{2}+a^{2}}\ .
\label{eq:acel_grav_ext_Kerr-Newman-Kasuya}
\end{equation}
%
%
\subsection{Thermodynamics}
Now, let us derive the interesting thermodynamic quantities. Firstly, let us take into account the Hawking radiation temperature in geometric units, $T_{+}$, and the surface area of the horizon in the presence of a cosmic string, $\mathcal{A}_{+,b}$, which are given by
\begin{equation}
T_{+}=\frac{\kappa_{+}}{2\pi}=\frac{1}{4\pi}\frac{r_{+}-r_{-}}{r_{+}^{2}+a^{2}}\ ,
\label{eq:temp_Hawking_Kerr-Newman-Kasuya}
\end{equation}
\begin{equation}
\mathcal{A}_{+,b}=\left.\int\int\sqrt{-g}\ d\theta\ d\phi \right|_{r=r_{+}}=4 \pi (r_{+}^{2}+a^{2})b\ ,
\label{eq:area_Kerr-Newman-Kasuya}
\end{equation}
where $g \equiv \mbox{det}(g_{\sigma\tau})=-b^{2}r^{4}\sin^{2}\theta$. Hence, the entropy at the event horizon, $S_{+,b}$, is given by
\begin{equation}
S_{+,b}=\frac{\mathcal{A}_{+,b}}{4}=\pi (r_{+}^{2}+a^{2})b\ ,
\label{eq:entropia_Kerr-Newman-Kasuya}
\end{equation}
where the subscript ``$b$'' codifies the presence of the cosmic string.

From Eq.~(\ref{eq:metrica_Kerr-Newman-Kasuya_string}), the dragging angular velocity of the exterior horizon, $\Omega_{+,b}$, is given by
\begin{equation}
\Omega_{+,b}=\left.-\frac{g_{03}}{g_{33}}\right|_{r=r_{+}}=\frac{a}{(r_{+}^{2}+a^{2})b}\ .
\label{eq:vel_ang_Kerr-Newman-Kasuya}
\end{equation}
The electric potential, $\Phi_{+}$, very near the event horizon, $r_{+}$, is
\begin{equation}
\Phi_{+}=\frac{Q_{e}r_{+}-\xi Q_{m}a}{r_{+}^{2}+a^{2}}\ ,
\label{eq:pot_ele_Kerr-Newman-Kasuya}
\end{equation}
where $\xi=\pm 1$. These two signs correspond to the two gauges which we will be using and means: the upper puts the string along the negative $z$ axis ($\theta=\pi$); the lower puts it along the positive $z$ axis ($\theta=0$).

The Komar's mass and angular momentum are expressed as
\begin{equation}
M_{phys}=Mb\ ,
\label{eq:massa_Kerr-Newman-Kasuya}
\end{equation}
\begin{equation}
J_{phys}=Jb\ ,
\label{eq:mom_ang_Kerr-Newman-Kasuya}
\end{equation}
so that the ratio $a=J/M=J_{phys}/M_{phys}$ remains unchanged. From Eq.~(\ref{eq:massa_Kerr-Newman-Kasuya}) we obtain
\begin{equation}
Eb=E_{phys}\ ,
\label{eq:energia_Kerr-Newman-Kasuya}
\end{equation}
which means that the physical energy of the black hole decreases due to the presence of the cosmic string.

These quantities given by Eqs.~(\ref{eq:temp_Hawking_Kerr-Newman-Kasuya})-(\ref{eq:energia_Kerr-Newman-Kasuya}) for the event horizon satisfy the first law of thermodynamics, namely,
\begin{equation}
dE_{phys}=T_{+}\ dS_{+,b}+\Omega_{+,b}\ dJ_{phys}+\Phi_{+}\ dQ_{e}\ .
\label{eq:1_lei_termo_Kerr-Newman-Kasuya}
\end{equation}
%
%
\section{Separation of the Klein-Gordon equation}
The covariant Klein-Gordon equation for a charged massive scalar field in the spacetime considered given by Eq.~(\ref{eq:metrica_Kerr-Newman-Kasuya_string}) can be written as
\begin{eqnarray}
&& \biggl\{-\frac{(r^{2}+a^{2})^{2}-\Delta a^{2}\sin^{2}\theta}{\Delta\rho^{2}}\frac{\partial^{2}}{\partial t^{2}}+\frac{1}{\rho^{2}}\frac{\partial}{\partial r}\left(\Delta\frac{\partial}{\partial r}\right)\nonumber\\
& + &\frac{1}{\rho^{2}\sin\theta}\frac{\partial}{\partial\theta}\left(\sin\theta\frac{\partial}{\partial\theta}\right)+\frac{\Delta -a^{2}\sin^{2}\theta}{b^{2}\Delta\rho^{2}\sin ^{2}\theta}\frac{\partial^{2}}{\partial\phi^{2}}\nonumber\\
& - &\frac{2a[(r^{2}+a^{2})-\Delta]}{b\Delta\rho^{2}}\frac{\partial^{2}}{\partial t\ \partial\phi}\nonumber\\
& - &2ie\frac{1}{\Delta\rho^{2}}\left\{Q_{e}r(r^{2}+a^{2})-Q_{m}a[\Delta\cos\theta-\xi\Delta+\xi(r^{2}+a^{2})]\right\}\frac{\partial}{\partial t}\nonumber\\
& - &2ie\frac{1}{b\Delta\rho^{2}}\left[Q_{e}ra-Q_{m}\left(\frac{\Delta\cos\theta}{\sin^{2}\theta}-\xi\frac{\Delta}{\sin^{2}\theta}+\xi a^{2}\right)\right]\frac{\partial}{\partial\phi}\nonumber\\
& - &e^{2}\frac{1}{\Delta\rho^{2}}\left\{-(Q_{e}r-\xi Q_{m}a)^{2}-Q_{m}^{2}\Delta\left[1-\frac{2(1-\xi\cos\theta)}{\sin^{2}\theta}\right]\right\}\nonumber\\
& - &\mu_{0}^{2}\biggr\}\Psi=0\ ,
\label{eq:mov_Kerr-Newman-Kasuya}
\end{eqnarray}
where $\mu_{0}$ is the mass of the scalar particle, $e$ is the charge of the particle, and the 4-vector electromagnetic potential given by \cite{GenRelativGrav.43.833}
\begin{eqnarray}
A_{\sigma}dx^{\sigma} & = & -\left(Q_{e}\frac{r}{\rho^{2}}-Q_{m}\frac{a\cos\theta}{\rho^{2}}\right)dt\nonumber\\
& + & b\left[Q_{e}\frac{ar\sin^{2}\theta}{\rho^{2}}+Q_{m}\left(\xi-\frac{r^{2}+a^{2}}{\rho^{2}}\cos\theta\right)\right]d\phi\ ,
\label{eq:potencial_EM_Kerr-Newman-Kasuya}
\end{eqnarray}
were used. Note that the units $G \equiv c \equiv \hbar \equiv k_{B} \equiv 1$ were chosen.

Due to the time independence and symmetry of the spacetime, the solution of Eq.~(\ref{eq:mov_Kerr-Newman-Kasuya}) can be written as
\begin{equation}
\Psi=\Psi(\mathbf{r},t)=R(r)S(\theta)\mbox{e}^{i(m+\tilde{\xi}eQ_{m})\phi}\mbox{e}^{-i\omega t}\ ,
\label{eq:separacao_variaveis_Kerr-Newman-Kasuya}
\end{equation}
where $\tilde{\xi}=(\xi,0)$, $m=\pm 1,\pm 2,\pm 3,...$ is the azimuthal quantum number, and the energy (frequency) is taken as $\omega > 0$ corresponding to a flux of particles at infinity.

Substituting Eq.~(\ref{eq:separacao_variaveis_Kerr-Newman-Kasuya}) into (\ref{eq:mov_Kerr-Newman-Kasuya}), we find that
\begin{eqnarray}
&& \frac{(r^{2}+a^{2})^{2}\omega^{2}}{\Delta}-a^{2}\omega^{2}\sin^{2}\theta+\frac{1}{R}\frac{d}{dr}\left(\Delta\frac{dR}{dr}\right)+\frac{1}{S}\frac{1}{\sin\theta}\frac{d}{d\theta}\left(\sin\theta\frac{dS}{d\theta}\right)\nonumber\\
&& -\frac{(m+\tilde{\xi}eQ_{m})^{2}}{b^{2}\sin^{2}\theta}+\frac{a^{2}(m+\tilde{\xi}eQ_{m})^{2}}{b^{2}\Delta}-\frac{2a\omega(r^{2}+a^{2})(m+\tilde{\xi}eQ_{m})}{b\Delta}\nonumber\\
&& +\frac{2a\omega(m+\tilde{\xi}eQ_{m})}{b}-\frac{2e\omega(r^{2}+a^{2})(Q_{e}r-\xi Q_{m}a)}{\Delta}+2e\omega Q_{m}a(\cos\theta-\xi)\nonumber\\
&& +\frac{2e(Q_{e}r-\xi Q_{m}a)a(m+\tilde{\xi}eQ_{m})}{b\Delta}-\frac{2eQ_{m}(\cos\theta-\xi)(m+\tilde{\xi}eQ_{m})}{b\sin^{2}\theta}\nonumber\\
&& +\frac{e^{2}(Q_{e}r-\xi Q_{m}a)^{2}}{\Delta}+e^{2}Q_{m}^{2}-\frac{2e^{2}Q_{m}^{2}(1-\xi\cos\theta)}{\sin^{2}\theta}\nonumber\\
&& -\mu_{0}^{2}r^{2}-\mu_{0}^{2}a^{2}\cos^{2}\theta=0\ .
\label{eq:mov_separavel_Kerr-Newman-Kasuya}
\end{eqnarray}

In what follows, we will separate the angular and radial parts of Eq.~(\ref{eq:mov_separavel_Kerr-Newman-Kasuya}) into two cases, namely, $\tilde{\xi}=\xi$ and $\tilde{\xi}=0$. In doing so, we can discuss more  appropriately, which is the most suitable separation to study the Hawking radiation.
%
%
\subsection{Case 1: Take \texorpdfstring{$\tilde{\xi}=\xi$}{tilde(xi)=xi} and consider the substitution \texorpdfstring{$\sin^{2}\theta=1-\cos^{2}\theta$}{(sinT)2=1-(cosT)2}}
In this case, Eq.(\ref{eq:mov_separavel_Kerr-Newman-Kasuya}) can be separated into the following equations for $S(\theta)$ and $R(r)$:
\begin{eqnarray}
&& \frac{1}{\sin\theta}\frac{d}{d\theta}\left(\sin\theta\frac{dS}{d\theta}\right)+\biggl\{-\frac{1}{\sin^{2}\theta}\left[m_{(b)}+eQ_{m}\cos\theta+eQ_{m}\xi f(b)\right]^{2}\nonumber\\
& + & c_{0}^{2}\cos^{2}\theta+2a\omega eQ_{m}\left[\cos\theta+\xi f(b)\right]+\lambda\biggr\}S=0
\label{eq:mov_angular_Kerr-Newman-Kasuya_case1}
\end{eqnarray}
and
\begin{eqnarray}
&& \frac{d}{dr}\left(\Delta\frac{dR}{dr}\right)+\biggl\{\frac{1}{\Delta}\left[(r^{2}+a^{2})\omega-am_{(b)}-eQ_{e}r-aeQ_{m}\xi f(b)\right]^{2}\nonumber\\
& - & \mu_{0}^{2}r^{2}+2a\omega m_{(b)}-a^{2}\omega^{2}-\lambda\biggr\}R=0\ ,
\label{eq:mov_radial_Kerr-Newman-Kasuya_case1}
\end{eqnarray}
where $\lambda$ is constant, $c_{0}^{2}=a^{2}(\omega^{2}-\mu_{0}^{2})$, $m_{(b)}=m/b$, and $f(b)=1/b-1$.

Note that in the $e=0$ case, Eqs.~(\ref{eq:mov_angular_Kerr-Newman-Kasuya_case1}) and (\ref{eq:mov_radial_Kerr-Newman-Kasuya_case1}) reduce to Eqs.~(7) and (8), respectively, of our paper \cite{ClassQuantumGrav.31.045003}, if we take $b=1$. In this separation, we have used the identity $\sin^{2}\theta=1-\cos^{2}\theta$ in the second term of Eq.~(\ref{eq:mov_separavel_Kerr-Newman-Kasuya}). However, as can be seen in \cite{ClassQuantumGrav.31.045003}, this separation gives us solutions of the radial equation in terms of the confluent Heun functions, in which its parameters contain terms proportional to $M$, the mass of the black hole, which is a result not suitable to study the Hawking radiation.

In that limit $b=1$ and $e \rightarrow -q$, Eqs.~(\ref{eq:mov_angular_Kerr-Newman-Kasuya_case1}) and (\ref{eq:mov_radial_Kerr-Newman-Kasuya_case1}) reduce to Eqs.~(9) and (10), respectively, presented in Semiz's paper \cite{PhysRevD.45.532}.
%
%
\subsection{Case 2: Take \texorpdfstring{$\tilde{\xi}=0$}{xi=0} and maintain \texorpdfstring{$\sin^{2}\theta$}{(sinT)2}}
In this case, we will maintain $\sin^{2}\theta$ in the second term of Eq.~(\ref{eq:mov_separavel_Kerr-Newman-Kasuya}), instead of substituting by $1-\cos^{2}\theta$. Thus, we can separate this equation according to
\begin{eqnarray}
&& \frac{1}{\sin\theta}\frac{d}{d\theta}\left(\sin\theta\frac{dS}{d\theta}\right)+\biggl\{-\left[\omega a\sin\theta-\frac{m_{(b)}+eQ_{m}(\cos\theta-\xi)}{\sin\theta}\right]^{2}\nonumber\\
& - & \mu_{0}^{2}a^{2}\cos^{2}\theta+\lambda\biggr\}S=0
\label{eq:mov_angular_Kerr-Newman-Kasuya_case2}
\end{eqnarray}
and
\begin{eqnarray}
&& \frac{d}{dr}\left(\Delta\frac{dR}{dr}\right)+\biggl\{\frac{1}{\Delta}\left[(r^{2}+a^{2})\omega-am_{(b)}-e(Q_{e}r-\xi Q_{m}a)\right]^{2}\nonumber\\
& - & \left(\mu_{0}^{2}r^{2}+\lambda\right)\biggr\}R=0\ .
\label{eq:mov_radial_Kerr-Newman-Kasuya_case2}
\end{eqnarray}
Note that in the $Q_{m}=0$ case, Eqs.~(\ref{eq:mov_angular_Kerr-Newman-Kasuya_case2}) and (\ref{eq:mov_radial_Kerr-Newman-Kasuya_case2}) reduce to Eqs.~(24) and (25), respectively, of our paper \cite{AnnPhys.350.14}, if we take $b=1$. Now, as can be seen in \cite{AnnPhys.350.14}, this separation gives us solutions of the radial equation in terms of the confluent Heun functions in a form suitable to study the Hawking radiation.
%
%
\section{Exact solutions of the Klein-Gordon equation}
In what follows we will solve the angular and radial parts of the Klein-Gordon equation for both cases above discussed.
%
%
\subsection{Case 1}
\subsubsection{Angular equation}\label{Angular_equation}
In order to obtain the solution of the angular part given by Eq.~(\ref{eq:mov_angular_Kerr-Newman-Kasuya_case1}), let us perform a change of variable, defining a new angular coordinate, $z$, such that
\begin{equation}
z=\frac{\cos\theta+1}{2}\ .
\label{eq:coord_angular_Kerr-Newman-Kasuya_case1_z}
\end{equation}
With this transformation, Eq.~(\ref{eq:mov_angular_Kerr-Newman-Kasuya_case1}) turns into
\begin{eqnarray}
&& \frac{d^{2}S}{dz^{2}}+\left(\frac{1}{z}+\frac{1}{z-1}\right)\frac{dS}{dz}+\biggl\{-(2c_{0})^{2}\nonumber\\
&& +\frac{4 a e Q_{m} \omega  [\xi f(b) -1]+2c_{0}^{2}+e^2 Q_{m}^2-[e \xi  Q_{m} f(b)+m_{(b)}]^2+2 \lambda }{2}\frac{1}{z}\nonumber\\
&& +\frac{-4 a e Q_{m} \omega  [\xi f(b) +1]-2 c_{0}^{2}-e^2 Q_{m}^2+[e \xi  Q_{m} f(b)+m_{(b)}]^2-2 \lambda}{2}\frac{1}{z-1}\nonumber\\
&& -\left[\frac{e \xi  Q_{m} f(b)-e Q_{m}+m_{(b)}}{2}\right]^2\frac{1}{z^2}\nonumber\\
&& -\left[\frac{e \xi  Q_{m} f(b)+e Q_{m}+m_{(b)}}{2}\right]^2\left.\frac{1}{(z-1)^2}\right\}S=0\ .
\label{eq:mov_angular_Kerr-Newman-Kasuya_case1_z}
\end{eqnarray}

Now, let us perform a transformation in order to eliminate the constant term and reduce the power of the terms proportional to $1/z^{2}$ and $1/(z-1)^{2}$. This transformation is a special case of the \textit{s-homotopic transformation} of the dependent variable, $S(z) \mapsto U(z)$, such that
\begin{equation}
S(z)=\mbox{e}^{A_{1}z}z^{A_{2}}(z-1)^{A_{3}}U(z)\ ,
\label{eq:s-homotopic_mov_angular_Kerr-Newman-Kasuya_case1_z}
\end{equation}
where the coefficients $A_{1}$, $A_{2}$, and $A_{3}$ are given by:
\begin{equation}
A_{1}=2c_{0}\ ;
\label{eq:A1_angular_Kerr-Newman-Kasuya_case1_z}
\end{equation}
\begin{equation}
A_{2}=\frac{e \xi  Q_{m} f(b)-e Q_{m}+m_{(b)}}{2}\ ;
\label{eq:A2_angular_Kerr-Newman-Kasuya_case1_z}
\end{equation}
\begin{equation}
A_{3}=\frac{e \xi  Q_{m} f(b)+e Q_{m}+m_{(b)}}{2}\ .
\label{eq:A3_angular_Kerr-Newman-Kasuya_case1_z}
\end{equation}
The function $U(z)$ satisfies the following equation
\begin{eqnarray}
&& \frac{d^{2}U}{dz^{2}}+\left(2 A_{1} +\frac{2 A_{2} +1}{z}+\frac{2 A_{3} +1}{z-1}\right)\frac{dU}{dz}\nonumber\\
&& +\left(\frac{2 A_{1} A_{2}+A_{1}-2 A_{2} A_{3}-A_{2}-A_{3}+A_{4}}{z}\right.\nonumber\\
&& +\left.\frac{2 A_{1} A_{3}+A_{1}+2 A_{2} A_{3}+A_{2}+A_{3}+A_{5}}{z-1}\right)U=0\ ,
\label{eq:mov_angular_Kerr-Newman-Kasuya_case1_z_U}
\end{eqnarray}
where the coefficients $A_{4}$, and $A_{5}$ are given by:
\begin{equation}
A_{4}=\frac{4 a e Q_{m} \omega  [\xi f(b) -1]+2c_{0}^{2}+e^2 Q_{m}^2-[e \xi  Q_{m} f(b)+m_{(b)}]^2+2 \lambda }{2}\ ;
\label{eq:A4_angular_Kerr-Newman-Kasuya_case1_z}
\end{equation}
\begin{equation}
A_{5}=\frac{-4 a e Q_{m} \omega  [\xi f(b) +1]-2 c_{0}^{2}-e^2 Q_{m}^2+[e \xi  Q_{m} f(b)+m_{(b)}]^2-2 \lambda}{2}\ .
\label{eq:A5_angular_Kerr-Newman-Kasuya_case1_z}
\end{equation}
Equation (\ref{eq:mov_angular_Kerr-Newman-Kasuya_case1_z_U}) is similar to the confluent Heun equation \cite{JPhysAMathTheor.43.035203} given by
\begin{equation}
\frac{d^{2}U}{dz^{2}}+\left(\alpha+\frac{\beta+1}{z}+\frac{\gamma+1}{z-1}\right)\frac{dU}{dz}+\left(\frac{\mu}{z}+\frac{\nu}{z-1}\right)U=0\ ,
\label{eq:Heun_confluente_forma_canonica}
\end{equation}
where $U(z)=\mbox{HeunC}(\alpha,\beta,\gamma,\delta,\eta;z)$ are the confluent Heun functions, with the parameters $\alpha$, $\beta$, $\gamma$, $\delta$, and $\eta$, related to $\mu$ and $\nu$ by
\begin{equation}
\mu=\frac{1}{2}(\alpha-\beta-\gamma+\alpha\beta-\beta\gamma)-\eta\ ,
\label{eq:mu_Heun_conlfuente_2}
\end{equation}
\begin{equation}
\nu=\frac{1}{2}(\alpha+\beta+\gamma+\alpha\gamma+\beta\gamma)+\delta+\eta\ ,
\label{eq:nu_Heun_conlfuente_2}
\end{equation}
according to the standard package of the \textbf{Maple}\texttrademark \textbf{17}.

Thus, the general solution of the angular part of the Klein-Gordon equation for a charged massive scalar field in the Kerr-Newman-Kasuya spacetime with a cosmic string, in the exterior region of the event horizon, given by Eq.~(\ref{eq:mov_angular_Kerr-Newman-Kasuya_case1_z}), over the entire range $0 \leq z < \infty$, can be written as
\begin{eqnarray}
S(z) & = & \mbox{e}^{\frac{1}{2}\alpha z}z^{\frac{1}{2}\beta}(z-1)^{\frac{1}{2}\gamma}\nonumber\\
& \times & \{C_{1}\ \mbox{HeunC}(\alpha,\beta,\gamma,\delta,\eta;z)\nonumber\\
& + & C_{2}\ z^{-\beta}\ \mbox{HeunC}(\alpha,-\beta,\gamma,\delta,\eta;z)\}\ ,
\label{eq:solucao_geral_angular_Kerr-Newman-Kasuya_case1_z}
\end{eqnarray}
where $C_{1}$ and $C_{2}$ are constants, and the parameters $\alpha$, $\beta$, $\gamma$, $\delta$, and $\eta$ are now given by:
\begin{equation}
\alpha=4 a (\omega ^2-\mu_{0}^{2})^{\frac{1}{2}}\ ;
\label{eq:alpha_angular_Kerr-Newman-Kasuya_case1_z}
\end{equation}
\begin{equation}
\beta=m_{(b)}+e Q_{m} [\xi f(b)-1]\ ;
\label{eq:beta_angular_Kerr-Newman-Kasuya_case1_z}
\end{equation}
\begin{equation}
\gamma=m_{(b)}+e Q_{m} [\xi f(b)+1]\ ;
\label{eq:gamma_angular_Kerr-Newman-Kasuya_case1_z}
\end{equation}
\begin{equation}
\delta=-4 a \omega e Q_{m}\ ;
\label{eq:delta_angular_Kerr-Newman-Kasuya_case1_z}
\end{equation}
\begin{equation}
\eta=-\frac{4 a e Q_{m} \omega  [\xi f(b) -1]+2c_{0}^{2}+e^2 Q_{m}^2-[e \xi  Q_{m} f(b)+m_{(b)}]^2+2 \lambda }{2}\ .
\label{eq:eta_angular_Kerr-Newman-Kasuya_case1_z}
\end{equation}
These two functions form linearly independent solutions of the confluent Heun dif\-fer\-en\-tial equation provided $\beta$ is not integer.
%
%
\subsubsection{Radial equation}\label{Radial_equation}
Now, in order to obtain the solution of the radial part given by Eq.~(\ref{eq:mov_radial_Kerr-Newman-Kasuya_case1}), let us use Eq.~(\ref{eq:superficie_hor_Kerr-Newman-Kasuya}).

With this definition, Eq.~(\ref{eq:mov_radial_Kerr-Newman-Kasuya_case1}) turns into
\begin{eqnarray}
&& \frac{d^{2}R}{dr^{2}}+\left(\frac{1}{r-r_{+}}+\frac{1}{r-r_{-}}\right)\frac{dR}{dr}\nonumber\\
& + & \frac{1}{(r-r_{+})(r-r_{-})}\biggl\{- (\mu_{0} ^2-\omega ^2)r^2- \omega  [2 e Q_{e}-\omega (r_{+}  +r_{-}) ]r\nonumber\\
& + & \omega ^2 (a^2+r_{+}^2+r_{+} r_{-}+r_{-}^2)-2 \omega e  [\xi Q_{m} a f(b)  +Q_{e} (r_{+}+r_{-})]+e^2 Q_{e}^2-\lambda\nonumber\\
& + & \frac{\{\omega (r_{+}^2+a^2) -a m_{(b)}-e[ Q_{e} r_{+}+\xi Q_{m} a f(b) ]\}^2}{(r_{+}-r_{-}) (r-r_{+}) }\nonumber\\
& + & \frac{\{\omega (r_{-}^2+a^2) -a m_{(b)}-e[ Q_{e} r_{-}+\xi Q_{m} a f(b) ]\}^2}{(r_{-}-r_{+}) (r-r_{-}) }\biggr\}R=0\ .
\label{eq:mov_radial_Kerr-Newman-Kasuya_case1_r}
\end{eqnarray}
This equation has singularities at $r=(a_{1},a_{2})=(r_{+},r_{-})$, and at $r=\infty$. The transformation of Eq.~(\ref{eq:mov_radial_Kerr-Newman-Kasuya_case1_r}) to a Heun-type equation is achieved by setting
\begin{equation}
x=\frac{r-a_{1}}{a_{2}-a_{1}}=\frac{r-r_{+}}{r_{-}-r_{+}}\ .
\label{eq:homog_subs_radial_Kerr-Newman-Kasuya_case1_x}
\end{equation}
Thus, we can write Eq.~(\ref{eq:mov_radial_Kerr-Newman-Kasuya_case1_r}) as
\begin{eqnarray}
&& \frac{d^{2}R}{dx^{2}}+\left(\frac{1}{x}+\frac{1}{x-1}\right)\frac{dR}{dx}+\biggl\{-[(\mu_{0} ^2-\omega ^2)^{\frac{1}{2}} (r_{+}-r_{-})]^2\nonumber\\
& + & \frac{2 a^{2}\{a \omega-[e \xi Q_{m} f(b)+m_{(b)}]\}^{2} +a^2 \{\omega ^2 (r_{+}+r_{-})^2-2 \omega e Q_{e}  (r_{+}+r_{-})\}}{(r_{+}-r_{-})^2 x}\nonumber\\
& + & \frac{(r_{+}-r_{-})^2 (\lambda +\mu_{0} ^2 r_{+}^2)-2 a \omega  [2 e \xi  Q_{m} f(b) r_{+} r_{-}+m_{(b)} (r_{+}^2+r_{-}^2)]}{(r_{+}-r_{-})^2 x}\nonumber\\
& + & \frac{2 e^2 Q_{e}^2 r_{+} r_{-}+2 e Q_{e} r_{+}^2 	\omega  (r_{+}-3 r_{-})-2 r_{+}^3 \omega ^2 (r_{+}-2 r_{-})}{(r_{+}-r_{-})^2 x}\nonumber\\
& + & \frac{2 a e Q_{e} (r_{+}+r_{-}) [e \xi  Q_{m} f(b)+m_{(b)}]}{(r_{+}-r_{-})^2 x}\nonumber\\
& - & \frac{2 a^{2}\{a \omega-[e \xi Q_{m} f(b)+m_{(b)}]\}^{2} +a^2 \{\omega ^2 (r_{+}+r_{-})^2-2 \omega e Q_{e}  (r_{+}+r_{-})\}}{(r_{+}-r_{-})^2 (x-1)}\nonumber\\
& - & \frac{(r_{+}-r_{-})^2 (\lambda +\mu_{0} ^2 r_{-}^2)-2 a \omega  [2 e \xi  Q_{m} f(b) r_{+} r_{-}+m_{(b)} (r_{+}^2+r_{-}^2)]}{(r_{+}-r_{-})^2 (x-1)}\nonumber\\
& - & \frac{2 e^2 Q_{e}^2 r_{+} r_{-}+2 e Q_{e} r_{-}^2 	\omega  (r_{-}-3 r_{+})-2 r_{-}^3 \omega ^2 (r_{-}-2 r_{+})}{(r_{+}-r_{-})^2 (x-1)}\nonumber\\
& - & \frac{2 a e Q_{e} (r_{+}+r_{-}) [e \xi  Q_{m} f(b)+m_{(b)}]}{(r_{+}-r_{-})^2 (x-1)}\nonumber\\
& - & \left\{i\frac{ \omega(r_{+}^2+a^2) -a m_{(b)} - e [ Q_{e} r_{+} + \xi Q_{m} a f(b) ]  }{ r_{+}-r_{-}}\right\}^{2}\frac{1}{x^2}\nonumber\\
& - & \left\{i\frac{ \omega(r_{-}^2+a^2) -a m_{(b)} - e [ Q_{e} r_{-} + \xi Q_{m} a f(b) ]  }{ r_{+}-r_{-}}\right\}^{2}\frac{1}{(x-1)^2}\biggr\}R=0\ .
\label{eq:mov_radial_Kerr-Newman-Kasuya_case1_x}
\end{eqnarray}

Now, let us perform a transformation in order to eliminate the constant term and reduce the power of the terms proportional to $1/x^{2}$ and $1/(x-1)^{2}$. Analogously to the previous situation, this transformation is a special case of the \textit{s-homotopic transformation} of the dependent variable, $R(x) \mapsto U(x)$, such that
\begin{equation}
R(x)=\mbox{e}^{B_{1}x}x^{B_{2}}(x-1)^{B_{3}}U(x)\ ,
\label{eq:s-homotopic_mov_radial_Kerr-Newman-Kasuya_case1_x}
\end{equation}
where the coefficients $B_{1}$, $B_{2}$, and $B_{3}$ are given by:
\begin{equation}
B_{1}=(\mu_{0} ^2-\omega ^2)^{\frac{1}{2}} (r_{+}-r_{-})\ ;
\label{eq:B1_radial_Kerr-Newman-Kasuya_case1_x}
\end{equation}
\begin{equation}
B_{2}=i\frac{ \omega(r_{+}^2+a^2) -a m_{(b)} - e [ Q_{e} r_{+} + \xi Q_{m} a f(b) ]  }{ r_{+}-r_{-}}\ ;
\label{eq:B2_radial_Kerr-Newman-Kasuya_case1_x}
\end{equation}
\begin{equation}
B_{3}=i\frac{ \omega(r_{-}^2+a^2) -a m_{(b)} - e [ Q_{e} r_{-} + \xi Q_{m} a f(b) ]  }{ r_{+}-r_{-}}\ .
\label{eq:B3_radial_Kerr-Newman-Kasuya_case1_x}
\end{equation}
The function $U(x)$ satisfies the following equation
\begin{eqnarray}
&& \frac{d^{2}U}{dx^{2}}+\left(2 B_{1} +\frac{2 B_{2} +1}{x}+\frac{2 B_{3} +1}{x-1}\right)\frac{dU}{dx}\nonumber\\
&& +\left(\frac{2 B_{1} B_{2}+B_{1}-2 B_{2} B_{3}-B_{2}-B_{3}+B_{4}}{x}\right.\nonumber\\
&& +\left.\frac{2 B_{1} B_{3}+B_{1}+2 B_{2} B_{3}+B_{2}+B_{3}+B_{5}}{x-1}\right)U=0\ ,
\label{eq:mov_radial_Kerr-Newman-Kasuya_case1_x_U}
\end{eqnarray}
where the coefficients $B_{4}$, and $B_{5}$ are given by:
\begin{eqnarray}
&& B_{4}=\frac{2 a^{2}\{a \omega-[e \xi Q_{m} f(b)+m_{(b)}]\}^{2} +a^2 \omega ^2 (r_{+}+r_{-})^2}{(r_{+}-r_{-})^2 }\nonumber\\
&& +\frac{(r_{+}-r_{-})^2 (\lambda +\mu_{0} ^2 r_{+}^2)-2 a \omega  [2 e \xi  Q_{m} f(b) r_{+} r_{-}+m_{(b)} (r_{+}^2+r_{-}^2)]}{(r_{+}-r_{-})^2 }\nonumber\\
&& +\frac{2 e^2 Q_{e}^2 r_{+} r_{-}+2 e Q_{e} r_{+}^2 	\omega  (r_{+}-3 r_{-})-2 r_{+}^3 \omega ^2 (r_{+}-2 r_{-})}{(r_{+}-r_{-})^2 }\nonumber\\
&& +\frac{2 a e Q_{e} (r_{+}+r_{-}) [e \xi  Q_{m} f(b)+m_{(b)}]}{(r_{+}-r_{-})^2 }\nonumber\\
&& +\frac{-2 a^2 \omega e Q_{e}  (r_{+}+r_{-})\}}{(r_{+}-r_{-})^2 }\ ;
\label{eq:A4_radial_Kerr-Newman-Kasuya_case1_x}
\end{eqnarray}
\begin{eqnarray}
&& B_{5}=-\frac{2 a^{2}\{a \omega-[e \xi Q_{m} f(b)+m_{(b)}]\}^{2} +a^2 \omega ^2 (r_{+}+r_{-})^2}{(r_{+}-r_{-})^2 }\nonumber\\
&& -\frac{(r_{+}-r_{-})^2 (\lambda +\mu_{0} ^2 r_{-}^2)-2 a \omega  [2 e \xi  Q_{m} f(b) r_{+} r_{-}+m_{(b)} (r_{+}^2+r_{-}^2)]}{(r_{+}-r_{-})^2 }\nonumber\\
&& -\frac{2 e^2 Q_{e}^2 r_{+} r_{-}+2 e Q_{e} r_{-}^2 	\omega  (r_{-}-3 r_{+})-2 r_{-}^3 \omega ^2 (r_{-}-2 r_{+})}{(r_{+}-r_{-})^2 }\nonumber\\
&& -\frac{2 a e Q_{e} (r_{+}+r_{-}) [e \xi  Q_{m} f(b)+m_{(b)}]}{(r_{+}-r_{-})^2 }\nonumber\\
&& -\frac{-2 a^2 \omega e Q_{e}  (r_{+}+r_{-})\}}{(r_{+}-r_{-})^2 }\ .
\label{eq:A5_radial_Kerr-Newman-Kasuya_case1_x}
\end{eqnarray}
Equation (\ref{eq:mov_radial_Kerr-Newman-Kasuya_case1_x_U}) is similar to the confluent Heun equation (\ref{eq:Heun_confluente_forma_canonica}). Thus, its general solution, in the exterior region of the event horizon, given by Eq.~(\ref{eq:mov_radial_Kerr-Newman-Kasuya_case1_x}), over the entire range $0 \leq x < \infty$, can be written as
\begin{eqnarray}
R(x) & = & \mbox{e}^{\frac{1}{2}\alpha x}x^{\frac{1}{2}\beta}(x-1)^{\frac{1}{2}\gamma}\nonumber\\
& \times & \{C_{1}\ \mbox{HeunC}(\alpha,\beta,\gamma,\delta,\eta;x)\nonumber\\
& + & C_{2}\ x^{-\beta}\ \mbox{HeunC}(\alpha,-\beta,\gamma,\delta,\eta;x)\}\ ,
\label{eq:solucao_geral_radial_Kerr-Newman-Kasuya_case1_x}
\end{eqnarray}
where $C_{1}$ and $C_{2}$ are constants, and the parameters $\alpha$, $\beta$, $\gamma$, $\delta$, and $\eta$ are now given by:
\begin{equation}
\alpha=2(\mu_{0} ^2-\omega ^2)^{\frac{1}{2}} (r_{+}-r_{-})\ ;
\label{eq:alpha_radial_Kerr-Newman-Kasuya_case1_x}
\end{equation}
\begin{equation}
\beta=2i\frac{ \omega(r_{+}^2+a^2) -a m_{(b)} - e [ Q_{e} r_{+} + \xi Q_{m} a f(b) ]  }{ r_{+}-r_{-}}\ ;
\label{eq:beta_radial_Kerr-Newman-Kasuya_case1_x}
\end{equation}
\begin{equation}
\gamma=2i\frac{ \omega(r_{-}^2+a^2) -a m_{(b)} - e [ Q_{e} r_{-} + \xi Q_{m} a f(b) ]  }{ r_{+}-r_{-}}\ ;
\label{eq:gamma_radial_Kerr-Newman-Kasuya_case1_x}
\end{equation}
\begin{equation}
\delta=[2 \omega e Q_{e} + (\mu_{0} ^2-2 \omega ^2)(r_{+}+r_{-})](r_{+}-r_{-})\ ;
\label{eq:delta_radial_Kerr-Newman-Kasuya_case1_x}
\end{equation}
\begin{eqnarray}
&& \eta=-\frac{2 a^{2}\{a \omega-[e \xi Q_{m} f(b)+m_{(b)}]\}^{2} +a^2 \{\omega ^2 (r_{+}+r_{-})^2\}}{(r_{+}-r_{-})^2 }\nonumber\\
&& -\frac{(r_{+}-r_{-})^2 (\lambda +\mu_{0} ^2 r_{+}^2)-2 a \omega  [2 e \xi  Q_{m} f(b) r_{+} r_{-}+m_{(b)} (r_{+}^2+r_{-}^2)]}{(r_{+}-r_{-})^2 }\nonumber\\
&& -\frac{2 e^2 Q_{e}^2 r_{+} r_{-}+2 e Q_{e} r_{+}^2 	\omega  (r_{+}-3 r_{-})-2 r_{+}^3 \omega ^2 (r_{+}-2 r_{-})}{(r_{+}-r_{-})^2 }\nonumber\\
&& -\frac{2 a e Q_{e} (r_{+}+r_{-}) [e \xi  Q_{m} f(b)+m_{(b)}]}{(r_{+}-r_{-})^2 }\nonumber\\
&& -\frac{-2 \omega e Q_{e}  (r_{+}+r_{-})\}}{(r_{+}-r_{-})^2 }\ ;
\label{eq:eta_radial_Kerr-Newman-Kasuya_case1_x}
\end{eqnarray}
These two functions form linearly independent solutions of the confluent Heun dif\-fer\-en\-tial equation provided $\beta$ is not integer. However, there is not any specific physical reason to impose that $\beta$ should be integer.

Note the dependence of both angular and radial solutions with the parameter $b$, associated with the presence of the cosmic string. If we take $b=1$, this implies that $f(b)=0$, so that the dependence of the parameters $\alpha$, $\beta$, $\gamma$, $\delta$, and $\eta$ relative to the signs which correspond to the two gauges, $\xi$, is eliminated. Thus, in this limit, and taking $e \rightarrow -q$, Eqs.~(\ref{eq:solucao_geral_angular_Kerr-Newman-Kasuya_case1_z}) and (\ref{eq:solucao_geral_radial_Kerr-Newman-Kasuya_case1_x}) are the general exact solutions for Eqs.~(9) and (10), respectively, of the Semiz's paper \cite{PhysRevD.45.532}.
%
%
\subsection{Case 2}
\subsubsection{Angular equation}
This case generalizes the results obtained in the paper by Fernandes et al. \cite{ClassQuantumGrav.23.7063} and our previous one \cite{EPL.109.60006}.

Following the same procedure of Section \ref{Angular_equation}, Eq.~(\ref{eq:mov_angular_Kerr-Newman-Kasuya_case2}) can be rewritten in a more convenient form, by defining a new angular coordinate, $z$, such that
\begin{equation}
z=\frac{\cos\theta+1}{2}\ ,
\label{eq:coord_angular_Kerr-Newman-Kasuya_case2_z}
\end{equation}
and performing a transformation in order to eliminate the constant term and reduce the power of the terms proportional to $1/z^{2}$ and $1/(z-1)^{2}$. This transformation is also a special case of the \textit{s-homotopic transformation} of the dependent variable, $S(z) \mapsto U(z)$, such that
\begin{equation}
S(z)=\mbox{e}^{A_{1}z}z^{A_{2}}(z-1)^{A_{3}}U(z)\ ,
\label{eq:s-homotopic_mov_angular_Kerr-Newman-Kasuya_case2_z}
\end{equation}
where the coefficients $A_{1}$, $A_{2}$, and $A_{3}$ are given by:
\begin{equation}
A_{1}=2a (\omega ^2- \mu_{0} ^2)^{\frac{1}{2}}\ ;
\label{eq:A1_angular_Kerr-Newman-Kasuya_case2_z}
\end{equation}
\begin{equation}
A_{2}=\frac{m_{(b)}-e Q_{m}(\xi+1)}{2}\ ;
\label{eq:A2_angular_Kerr-Newman-Kasuya_case2_z}
\end{equation}
\begin{equation}
A_{3}=\frac{m_{(b)}-e Q_{m}(\xi-1)}{2}\ .
\label{eq:A3_angular_Kerr-Newman-Kasuya_case2_z}
\end{equation}
The function $U(z)$ satisfies the following equation
\begin{eqnarray}
&& \frac{d^{2}U}{dz^{2}}+\left(2 A_{1} +\frac{2 A_{2} +1}{z}+\frac{2 A_{3} +1}{z-1}\right)\frac{dU}{dz}\nonumber\\
&& +\left(\frac{2 A_{1} A_{2}+A_{1}-2 A_{2} A_{3}-A_{2}-A_{3}+A_{4}}{z}\right.\nonumber\\
&& +\left.\frac{2 A_{1} A_{3}+A_{1}+2 A_{2} A_{3}+A_{2}+A_{3}+A_{5}}{z-1}\right)U=0\ ,
\label{eq:mov_angular_Kerr-Newman-Kasuya_case2_z_U}
\end{eqnarray}
where the coefficients $A_{4}$, and $A_{5}$ are given by:
\begin{equation}
A_{4}=\frac{-2 a^2 \mu_{0}^{2}+4 a \omega  [m_{(b)}-e Q_{m}(\xi +1)]+e^2 Q_{m}^{2}-(m_{(b)}-e \xi Q_{m})^2+2 \lambda }{2}\ ;
\label{eq:A4_angular_Kerr-Newman-Kasuya_case2_z}
\end{equation}
\begin{equation}
A_{5}=\frac{2 a^2 \mu_{0}^{2}-4 a \omega  [m_{(b)}-e Q_{m}(\xi -1)]-e^2 Q_{m}^{2}+(m_{(b)}-e \xi Q_{m})^2-2 \lambda }{2}\ .
\label{eq:A5_angular_Kerr-Newman-Kasuya_case2_z}
\end{equation}
Note that Eq.~(\ref{eq:mov_angular_Kerr-Newman-Kasuya_case2_z_U}) is similar to the confluent Heun equation (\ref{eq:Heun_confluente_forma_canonica}).

Thus, the general solution of the angular part of the Klein-Gordon equation for a charged massive scalar field in the Kerr-Newman-Kasuya spacetime with a cosmic string, in the exterior region of the event horizon, given by Eq.~(\ref{eq:mov_angular_Kerr-Newman-Kasuya_case2}) over the entire range $0 \leq z < \infty$, can be written as
\begin{eqnarray}
S(z) & = & \mbox{e}^{\frac{1}{2}\alpha z}z^{\frac{1}{2}\beta}(z-1)^{\frac{1}{2}\gamma}\nonumber\\
& \times & \{C_{1}\ \mbox{HeunC}(\alpha,\beta,\gamma,\delta,\eta;z)\nonumber\\
& + & C_{2}\ z^{-\beta}\ \mbox{HeunC}(\alpha,-\beta,\gamma,\delta,\eta;z)\}\ ,
\label{eq:solucao_geral_angular_Kerr-Newman-Kasuya_case2_z}
\end{eqnarray}
where $C_{1}$ and $C_{2}$ are constants, and the parameters $\alpha$, $\beta$, $\gamma$, $\delta$, and $\eta$ are now given by:
\begin{equation}
\alpha=4 a (\omega ^2-\mu_{0}^{2})^{\frac{1}{2}}\ ;
\label{eq:alpha_angular_Kerr-Newman-Kasuya_case2_z}
\end{equation}
\begin{equation}
\beta=m_{(b)}-e Q_{m}(\xi +1)\ ;
\label{eq:beta_angular_Kerr-Newman-Kasuya_case2_z}
\end{equation}
\begin{equation}
\gamma=m_{(b)}-e Q_{m}(\xi -1)\ ;
\label{eq:gamma_angular_Kerr-Newman-Kasuya_case2_z}
\end{equation}
\begin{equation}
\delta=-4 a \omega e Q_{m}\ ;
\label{eq:delta_angular_Kerr-Newman-Kasuya_case2_z}
\end{equation}
\begin{equation}
\eta=-\frac{-2 a^2 \mu_{0}^{2}+4 a \omega  [m_{(b)}-e Q_{m}(\xi +1)]+e^2 Q_{m}^{2}-(m_{(b)}-e \xi Q_{m})^2+2 \lambda }{2}\ .
\label{eq:eta_angular_Kerr-Newman-Kasuya_case2_z}
\end{equation}
These two functions form linearly independent solutions of the confluent Heun dif\-fer\-en\-tial equation provided $\beta$ is not integer.
%
%
\subsubsection{Radial equation}
Now, let us obtain the exact and general solution for the radial part given by Eq.~(\ref{eq:mov_radial_Kerr-Newman-Kasuya_case2}). Following the same procedure of Section \ref{Radial_equation} and using Eq.~(\ref{eq:superficie_hor_Kerr-Newman-Kasuya}), the transformation of (\ref{eq:mov_radial_Kerr-Newman-Kasuya_case2}) to a Heun-type equation is achieved by setting
\begin{equation}
x=\frac{r-a_{1}}{a_{2}-a_{1}}=\frac{r-r_{+}}{r_{-}-r_{+}}\ ,
\label{eq:homog_subs_radial_Kerr-Newman-Kasuya_case2_x}
\end{equation}
and performing a transformation in order to eliminate the term constant and reduce the power of the terms proportional to $1/x^{2}$ and $1/(x-1)^{2}$. This transformation is also a special case of the \textit{s-homotopic transformation} of the dependent variable, $R(x) \mapsto U(x)$, such that
\begin{equation}
R(x)=\mbox{e}^{B_{1}x}x^{B_{2}}(x-1)^{B_{3}}U(x)\ ,
\label{eq:s-homotopic_mov_radial_Kerr-Newman-Kasuya_case2_x}
\end{equation}
where the coefficients $B_{1}$, $B_{2}$, and $B_{3}$ are given by:
\begin{equation}
B_{1}=(\mu_{0} ^2-\omega ^2)^{\frac{1}{2}} (r_{+}-r_{-})\ ;
\label{eq:B1_radial_Kerr-Newman-Kasuya_case2_x}
\end{equation}
\begin{equation}
B_{2}=i\frac{ \omega(r_{+}^2+a^2) -a m_{(b)} - e ( Q_{e} r_{+} - \xi Q_{m} a ) }{ r_{+}-r_{-}}\ ;
\label{eq:B2_radial_Kerr-Newman-Kasuya_case2_x}
\end{equation}
\begin{equation}
B_{3}=i\frac{ \omega(r_{-}^2+a^2) -a m_{(b)} - e ( Q_{e} r_{-} - \xi Q_{m} a ) }{ r_{+}-r_{-}}\ .
\label{eq:B3_radial_Kerr-Newman-Kasuya_case2_x}
\end{equation}
The function $U(z)$ satisfies the following equation
\begin{eqnarray}
&& \frac{d^{2}U}{dx^{2}}+\left(2 B_{1} +\frac{2 B_{2} +1}{x}+\frac{2 B_{3} +1}{x-1}\right)\frac{dU}{dx}\nonumber\\
&& +\left(\frac{2 B_{1} B_{2}+B_{1}-2 B_{2} B_{3}-B_{2}-B_{3}+B_{4}}{x}\right.\nonumber\\
&& +\left.\frac{2 B_{1} B_{3}+B_{1}+2 B_{2} B_{3}+B_{2}+B_{3}+B_{5}}{x-1}\right)U=0\ ,
\label{eq:mov_radial_Kerr-Newman-Kasuya_case2_x_U}
\end{eqnarray}
where the coefficients $B_{4}$, and $B_{5}$ are given by:
\begin{eqnarray}
&& B_{4}=\frac{2 a^{2}[a \omega-(m_{(b)}-e \xi Q_{m})]^{2} +4a^2 \omega ^2 r_{+}r_{-}}{(r_{+}-r_{-})^2 }\nonumber\\
&& +\frac{(r_{+}-r_{-})^2 (\lambda +\mu_{0} ^2 r_{+}^2)-4 a \omega  (m_{(b)}- e \xi  Q_{m}) r_{+} r_{-}}{(r_{+}-r_{-})^2 }\nonumber\\
&& +\frac{2 e^2 Q_{e}^2 r_{+} r_{-}+2 e Q_{e} r_{+}^2 	\omega  (r_{+}-3 r_{-})-2 r_{+}^3 \omega ^2 (r_{+}-2 r_{-})}{(r_{+}-r_{-})^2 }\nonumber\\
&& +\frac{2 a e Q_{e} (r_{+}+r_{-}) (m_{(b)}-e \xi  Q_{m})}{(r_{+}-r_{-})^2 }\nonumber\\
&& +\frac{-2 a^2 \omega e Q_{e}  (r_{+}+r_{-})\}}{(r_{+}-r_{-})^2 }\ ;
\label{eq:A4_radial_Kerr-Newman-Kasuya_case2_x}
\end{eqnarray}
\begin{eqnarray}
&& B_{5}=-\frac{2 a^{2}[a \omega-(m_{(b)}-e \xi Q_{m})]^{2} +4a^2 \omega ^2 r_{+}r_{-}}{(r_{+}-r_{-})^2 }\nonumber\\
&& -\frac{(r_{+}-r_{-})^2 (\lambda +\mu_{0} ^2 r_{-}^2)-4 a \omega  (m_{(b)}- e \xi  Q_{m}) r_{+} r_{-}}{(r_{+}-r_{-})^2 }\nonumber\\
&& -\frac{2 e^2 Q_{e}^2 r_{+} r_{-}+2 e Q_{e} r_{-}^2 	\omega  (r_{-}-3 r_{+})-2 r_{-}^3 \omega ^2 (r_{-}-2 r_{+})}{(r_{+}-r_{-})^2 }\nonumber\\
&& -\frac{2 a e Q_{e} (r_{+}+r_{-}) (m_{(b)}-e \xi  Q_{m})}{(r_{+}-r_{-})^2 }\nonumber\\
&& -\frac{-2 a^2 \omega e Q_{e}  (r_{+}+r_{-})\}}{(r_{+}-r_{-})^2 }\ ;
\label{eq:A5_radial_Kerr-Newman-Kasuya_case2_x}
\end{eqnarray}
It is worth calling attention to the fact that Eq.~(\ref{eq:mov_radial_Kerr-Newman-Kasuya_case2_x_U}) is similar to the confluent Heun equation (\ref{eq:Heun_confluente_forma_canonica}). Thus, the general solution of Eq.~(\ref{eq:mov_radial_Kerr-Newman-Kasuya_case2}), in the exterior region of the event horizon, over the entire range $0 \leq x < \infty$, can be written as
\begin{eqnarray}
R(x) & = & \mbox{e}^{\frac{1}{2}\alpha x}x^{\frac{1}{2}\beta}(x-1)^{\frac{1}{2}\gamma}\nonumber\\
& \times & \{C_{1}\ \mbox{HeunC}(\alpha,\beta,\gamma,\delta,\eta;x)\nonumber\\
& + & C_{2}\ x^{-\beta}\ \mbox{HeunC}(\alpha,-\beta,\gamma,\delta,\eta;x)\}\ ,
\label{eq:solucao_geral_radial_Kerr-Newman-Kasuya_case2_x}
\end{eqnarray}
where $C_{1}$ and $C_{2}$ are constants, and the parameters $\alpha$, $\beta$, $\gamma$, $\delta$, and $\eta$ are now given by:
\begin{equation}
\alpha=2(\mu_{0} ^2-\omega ^2)^{\frac{1}{2}} (r_{+}-r_{-})\ ;
\label{eq:alpha_radial_Kerr-Newman-Kasuya_case2_x}
\end{equation}
\begin{equation}
\beta=2i\frac{ \omega(r_{+}^2+a^2) -a m_{(b)} - e ( Q_{e} r_{+} - \xi Q_{m} a ) }{ r_{+}-r_{-}}\ ;
\label{eq:beta_radial_Kerr-Newman-Kasuya_case2_x}
\end{equation}
\begin{equation}
\gamma=2i\frac{ \omega(r_{-}^2+a^2) -a m_{(b)} - e ( Q_{e} r_{-} - \xi Q_{m} a ) }{ r_{+}-r_{-}}\ ;
\label{eq:gamma_radial_Kerr-Newman-Kasuya_case2_x}
\end{equation}
\begin{equation}
\delta=[2 \omega e Q_{e} + (\mu_{0} ^2-2 \omega ^2)(r_{+}+r_{-})](r_{+}-r_{-})\ ;
\label{eq:delta_radial_Kerr-Newman-Kasuya_case2_x}
\end{equation}
\begin{eqnarray}
&& \eta=-\frac{2 a^{2}[a \omega-(m_{(b)}-e \xi Q_{m})]^{2} +4a^2 \omega ^2 r_{+}r_{-}}{(r_{+}-r_{-})^2 }\nonumber\\
&& -\frac{(r_{+}-r_{-})^2 (\lambda +\mu_{0} ^2 r_{+}^2)-4 a \omega  (m_{(b)}- e \xi  Q_{m}) r_{+} r_{-}}{(r_{+}-r_{-})^2 }\nonumber\\
&& -\frac{2 e^2 Q_{e}^2 r_{+} r_{-}+2 e Q_{e} r_{+}^2 	\omega  (r_{+}-3 r_{-})-2 r_{+}^3 \omega ^2 (r_{+}-2 r_{-})}{(r_{+}-r_{-})^2 }\nonumber\\
&& -\frac{2 a e Q_{e} (r_{+}+r_{-}) (m_{(b)}-e \xi  Q_{m})}{(r_{+}-r_{-})^2 }\nonumber\\
&& -\frac{-2 a^2 \omega e Q_{e}  (r_{+}+r_{-})\}}{(r_{+}-r_{-})^2 }\ ;
\label{eq:eta_radial_Kerr-Newman-Kasuya_case2_x}
\end{eqnarray}
These two functions form linearly independent solutions of the confluent Heun dif\-fer\-en\-tial equation provided $\beta$ is not integer. However, there is not any specific physical reason to impose that $\beta$ should be integer.

Note, again, the dependence of both angular and radial solutions with the parameter $b$, associated with the presence of the cosmic string.

If we consider the expansion in power series of the confluent Heun functions with respect to the independent variable $x$, in a vicinity of the regular singular point $x=0$ \cite{Ronveaux:1995}, we can write
\begin{eqnarray}
\mbox{HeunC}(\alpha,\beta,\gamma,\delta,\eta;x) & = & 1+\frac{1}{2}\frac{(-\alpha\beta+\beta\gamma+2\eta-\alpha+\beta+\gamma)}{(\beta+1)}x\nonumber\\
& + & \frac{1}{8}\frac{1}{(\beta+1)(\beta+2)}\left(\alpha^{2}\beta^{2}\right.-2\alpha\beta^{2}\gamma+\beta^{2}\gamma^{2}\nonumber\\
& - & 4\eta\alpha\beta+4\eta\beta\gamma+4\alpha^{2}\beta-2\alpha\beta^{2}-6\alpha\beta\gamma\nonumber\\
& + & 4\beta^{2}\gamma+4\beta\gamma^{2}+4\eta^{2}-8\eta\alpha+8\eta\beta+8\eta\gamma\nonumber\\
& + & 3\alpha^{2}-4\alpha\beta-4\alpha\gamma+3\beta^{2}+4\beta\delta\nonumber\\
& + & \left.10\beta\gamma+3\gamma^{2}+8\eta+4\beta+4\delta+4\gamma\right)x^2+...\ ,
\label{eq:serie_HeunC_todo_x}
\end{eqnarray}
which is a useful form to be used in the discussion of Hawking radiation.
%
%
\section{Black hole radiation and analytic continuation}
We will consider the charged massive scalar field near the horizon in order to discuss the Hawking radiation.

From Eqs.~(\ref{eq:homog_subs_radial_Kerr-Newman-Kasuya_case2_x}) and (\ref{eq:serie_HeunC_todo_x}) we can see that the radial solution given by Eq.~(\ref{eq:solucao_geral_radial_Kerr-Newman-Kasuya_case2_x}), near the exterior event horizon, that is, when $r \rightarrow r_{+}$ which implies that $x \rightarrow 0$, behaves asymptotically as
\begin{equation}
R(r) \sim C_{1}\ (r-r_{+})^{\beta/2}+C_{2}\ (r-r_{+})^{-\beta/2}\ ,
\label{eq:exp_0_solucao_geral_radial_Kerr-Newman-Kasuya_case2_x}
\end{equation}
where we are considering contributions only of the first term in the expansion, and all constants are included in $C_{1}$ and $C_{2}$. Thus, considering the time factor, near the black hole event horizon $r_{+}$, this solution is given by
\begin{equation}
\Psi=\mbox{e}^{-i \omega t}(r-r_{+})^{\pm\beta/2}\ .
\label{eq:sol_onda_radial_Kerr-Newman-Kasuya_case2_x}
\end{equation}
From Eq.~(\ref{eq:beta_radial_Kerr-Newman-Kasuya_case2_x}), we obtain
\begin{equation}
\frac{\beta}{2}=i\left[\omega\frac{r_{+}^{2}+a^{2}}{r_{+}-r_{-}}-\left(m\frac{a}{(r_{+}-r_{-})b}+e\frac{Q_{e} r_{+} - \xi Q_{m} a}{r_{+}-r_{-}}\right)\right]\ ,
\label{eq:beta/2_radial_Kerr-Newman-Kasuya_case2_x}
\end{equation}
which can be rewritten as
\begin{eqnarray}
\frac{\beta}{2} & = & \frac{i}{2\kappa_{+}}[\omega-(m\Omega_{+,b}+e\Phi_{+})]\nonumber\\
& = & \frac{i}{2\kappa_{+}}(\omega-\omega_{0,b})\ ,
\label{eq:expoente_rad_Hawking_Kerr-Newman-Kasuya_case2_x}
\end{eqnarray}
by using Eqs.~(\ref{eq:acel_grav_ext_Kerr-Newman-Kasuya}), (\ref{eq:vel_ang_Kerr-Newman-Kasuya}), and (\ref{eq:pot_ele_Kerr-Newman-Kasuya}), where $\omega_{0,b}=m\Omega_{+,b}+e\Phi_{+}$.

Therefore, on the black hole exterior horizon surface, the ingoing and outgoing wave solutions are
\begin{equation}
\Psi_{in}=\mbox{e}^{-i \omega t}(r-r_{+})^{-\frac{i}{2\kappa_{+}}(\omega-\omega_{0,b})}\ ,
\label{eq:sol_in_1_Kerr-Newman-Kasuya_case2_x}
\end{equation}
\begin{equation}
\Psi_{out}(r>r_{+})=\mbox{e}^{-i \omega t}(r-r_{+})^{\frac{i}{2\kappa_{+}}(\omega-\omega_{0,b})}\ .
\label{eq:sol_out_2_Kerr-Newman-Kasuya_case2_x}
\end{equation}
These solutions depend on the parameter $b$, in such a way that the total energy of the radiated particles is decreased due to presence of the cosmic string. These solutions for the scalar fields near the horizon will be useful to investigate Hawking radiation of charged massive scalar particles. It is worth calling attention to the fact that we are using the analytical solution of the radial part of the Klein-Gordon equation in the spacetime under consideration, differently from the calculation usually done in the literature, as for example in \cite{KexueTongbao.29.1303,ChinPhysLett.22.2492,ChinPhysB.17.2321}.

Using the definitions of the tortoise and Eddington-Finkelstein coordinates, we get:
\begin{equation}
\ln(r-r_{+})=\frac{1}{r_{+}^{2}+a^{2}}\left.\frac{d\Delta}{dr}\right|_{r=r_{+}}r_{*}=2\kappa_{+}r_{*}\ ;
\label{eq:coord_tortoise_1}
\end{equation}
\begin{equation}
\hat{r}=\frac{\omega-\omega_{0}}{\omega}r_{*}\ ;
\label{eq:hatr}
\end{equation}
\begin{equation}
v=t+\hat{r}
\label{eq:coord_Eddington-Finkelstein}\ .
\end{equation}
In those new coordinates, we obtain the following ingoing and outgoing wave solutions
\begin{eqnarray}
\Psi_{in} & = & \mbox{e}^{-i \omega v}\mbox{e}^{i \omega \hat{r}}(r-r_{+})^{-\frac{i}{2\kappa_{+}}(\omega-\omega_{0,b})}\nonumber\\
& = & \mbox{e}^{-i \omega v}\mbox{e}^{i (\omega-\omega_{0,b}) r_{*}}(r-r_{+})^{-\frac{i}{2\kappa_{+}}(\omega-\omega_{0,b})}\nonumber\\
& = & \mbox{e}^{-i \omega v}(r-r_{+})^{\frac{i}{2\kappa_{+}}(\omega-\omega_{0,b})}(r-r_{+})^{-\frac{i}{2\kappa_{+}}(\omega-\omega_{0,b})}\nonumber\\
& = & \mbox{e}^{-i \omega v}\ ,
\label{eq:sol_in_1_Kerr-Newman-Kasuya_case2_x_tortoise}
\end{eqnarray}
\begin{eqnarray}
\Psi_{out}(r>r_{+}) & = & \mbox{e}^{-i \omega v}\mbox{e}^{i \omega \hat{r}}(r-r_{+})^{\frac{i}{2\kappa_{+}}(\omega-\omega_{0,b})}\nonumber\\
& = & \mbox{e}^{-i \omega v}\mbox{e}^{i (\omega-\omega_{0,b}) r_{*}}(r-r_{+})^{\frac{i}{2\kappa_{+}}(\omega-\omega_{0,b})}\nonumber\\
& = & \mbox{e}^{-i \omega v}(r-r_{+})^{\frac{i}{2\kappa_{+}}(\omega-\omega_{0,b})}(r-r_{+})^{\frac{i}{2\kappa_{+}}(\omega-\omega_{0,b})}\nonumber\\
& = & \mbox{e}^{-i \omega v}(r-r_{+})^{\frac{i}{\kappa_{+}}(\omega-\omega_{0,b})}\ .
\label{eq:sol_out_2_Kerr-Newman-Kasuya_case2_x_tortoise}
\end{eqnarray}

The solutions (\ref{eq:sol_in_1_Kerr-Newman-Kasuya_case2_x_tortoise}) and (\ref{eq:sol_out_2_Kerr-Newman-Kasuya_case2_x_tortoise}), in the case $b=1$ and $Q_{m}=0$, are exactly the solutions obtained in our recent paper \cite{AnnPhys.350.14}. In the case $Q_{e}=Q_{m}=e=0$ and $a^{2} \approx 0$, the solutions (\ref{eq:sol_in_1_Kerr-Newman-Kasuya_case2_x_tortoise}) and (\ref{eq:sol_out_2_Kerr-Newman-Kasuya_case2_x_tortoise}) are exactly the solutions obtained in our most recent paper \cite{EPL.109.60006}.
%
%

Now, we will obtain by analytic continuation a real damped part of the outgoing wave solution of the scalar field which will be used to construct an explicit expression for the decay rate $\Gamma$. This real damped part corresponds (at least in part) to the temporal contribution to the decay rate \cite{ChinPhysLett.22.2492} found by the tunneling method used to investigate the Hawking radiation.

From Eq.~(\ref{eq:sol_out_2_Kerr-Newman-Kasuya_case2_x_tortoise}), we see that this solution is not analytical in the exterior event horizon, $r=r_{+}$. According to the Damour-Ruffini method \cite{PhysRevD.14.332}, by analytic continuation, rotating $-\pi$ through the lower-half complex $r$ plane, we obtain
\begin{equation}
(r-r_{+}) \rightarrow \left|r-r_{+}\right|\mbox{e}^{-i\pi}=(r_{+}-r)\mbox{e}^{-i\pi}\ .
\label{eq:rel_3_Kerr-Newman-Kasuya_case2_x}
\end{equation}
Thus, the outgoing wave solution on the horizon surface $r_{+}$ is
\begin{equation}
\Psi_{out}(r<r_{+})=\mbox{e}^{-i\omega v}(r_{+}-r)^{\frac{i}{\kappa_{+}}(\omega-\omega_{0,b})}\mbox{e}^{\frac{\pi}{\kappa_{+}}(\omega-\omega_{0,b})}\ .
\label{eq:sol_1_out_4_Kerr-Newman-Kasuya_case2_x}
\end{equation}
Equations (\ref{eq:sol_out_2_Kerr-Newman-Kasuya_case2_x_tortoise}) and (\ref{eq:sol_1_out_4_Kerr-Newman-Kasuya_case2_x}) describe the outgoing wave outside and inside of the black hole, respectively. Therefore, for an outgoing wave of a particle with energy $\omega$, charge $e$ and angular momentum $m$, the outgoing decay rate or the relative scattering probability of the scalar wave at the event horizon surface, $r=r_{+}$, is given by
\begin{equation}
\Gamma_{+}=\left|\frac{\Psi_{out}(r>r_{+})}{\Psi_{out}(r<r_{+})}\right|^{2}=\mbox{e}^{-\frac{2\pi}{\kappa_{+}}(\omega-\omega_{0,b})}\ ,
\label{eq:taxa_refl_Kerr-Newman-Kasuya_case2_x}
\end{equation}
This is the relative probability of creating a particle-antiparticle pair just outside the horizon.
%
%
\section{Bekenstein-Hawking entropy, black body spectrum and Hawking flux}
After the black hole event horizon radiates particles with energy $\omega$, charge $e$ and angular momentum $m$, in order to consider the reaction of the radiation of the particle to the spacetime, we must replace $M_{phys},Q_{e},J_{phys}$ by $M_{phys}-\omega,Q_{e}-e,J_{phys}-m$, respectively, in the line element of the spacetime under consideration. Doing these changes, we must guarantee that the total energy, charge and angular momentum of the spacetime are all conserved, that is,
\begin{equation}
\begin{array}{r}
-\omega=\triangle E_{phys}\ ,\\
-e=\triangle Q_{e}\ ,\\
-m=\triangle J_{phys}\ ,
\end{array}
\label{eq:param_cons_Kerr-Newman-Kasuya_case2_x}
\end{equation}
where $\triangle E_{phys}$, $\triangle Q_{e}$, and $\triangle J_{phys}$ are the physical energy, electric charge and Komar's angular momentum variations of the black hole event horizon, before and after the emission of radiation, respectively. Substituting Eqs.~(\ref{eq:1_lei_termo_Kerr-Newman-Kasuya}) and (\ref{eq:param_cons_Kerr-Newman-Kasuya_case2_x}) into Eq.(\ref{eq:taxa_refl_Kerr-Newman-Kasuya_case2_x}), we obtain the outgoing decay rate at the event horizon surface $r=r_{+}$:
\begin{eqnarray}
\Gamma_{+} & = & \mbox{e}^{-\frac{2\pi}{\kappa_{+}}(-\triangle E_{phys}-m\Omega_{+,b}-e\Phi_{+})}\nonumber\\
& = & \mbox{e}^{-\frac{2\pi}{\kappa_{+}}(-\triangle E_{phys}+\Omega_{+,b}\triangle J_{phys}+\Phi_{+}\triangle Q_{e})}\nonumber\\
& = & \mbox{e}^{-\frac{2\pi}{\kappa_{+}}(-T_{+}\triangle S_{+,b})}\nonumber\\
& = & \mbox{e}^{\triangle S_{+,b}}\ ,
\label{eq:taxa_refl_param_Kerr-Newman-Kasuya_case2_x}
\end{eqnarray}
where we have used Eq.~(\ref{eq:temp_Hawking_Kerr-Newman-Kasuya}). $\triangle S_{+,b}$ is the change of the Bekenstein-Hawking entropy in the presence of a cosmic string, if we compare situations before and after the emission of radiation. It is obtained from the expressions for the exterior event horizon (\ref{eq:sol_padrao_Kerr-Newman-Kasuya_1}) and for the entropy (\ref{eq:entropia_Kerr-Newman-Kasuya}), as follows:
\begin{eqnarray}
\triangle S_{+,b} & = & S_{+,b}(M_{phys}-\omega,Q_{e}-e,J_{phys}-m)-S_{+,b}(M_{phys},Q_{e},J_{phys})\nonumber\\
& = & \pi [r_{+}^{2}(M-\omega,Q_{e}-e,J-m)+a^{2}(M-\omega,J-m)]b\nonumber\\
& - & \pi [r_{+}^{2}(M,Q_{e},J)+a^{2}(M,J)]b\nonumber\\
& = & \pi[2(M-\omega)^{2}-(Q_{e}-e)^{2}-Q_{m}^{2}\nonumber\\
& + & 2(M-\omega)\sqrt{(M-\omega)^{2}-a_{\omega}^{2}-(Q_{e}-e)^{2}-Q_{m}^{2}}\nonumber\\
& - & 2M^{2}+Q_{e}^{2}+Q_{m}^{2}-2M\sqrt{M^{2}-a^{2}-Q_{e}^{2}-Q_{m}^{2}}]b\ ,
\label{eq:entropia_Bekenstein-Hawking_Kerr-Newman-Kasuya_case2_x}
\end{eqnarray}
where $a=a(M,J)=J/M$ and
\begin{equation}
a_{\omega}^{2}=\left(\frac{J-m}{M-\omega}\right)^{2}\ .
\label{eq:a_omega_Kerr-Newman}
\end{equation}

According to the Sannan heuristic derivation \cite{GenRelativGravit.20.239}, the mean number of particles emitted, $\bar{N}_{\omega}$, in a given mode can be obtained from the expression for the relative scattering probability, $\Gamma_{+}$, as follows:
\begin{equation}
\bar{N}_{\omega}=\frac{\Gamma_{+}}{1-\Gamma_{+}}=\frac{1}{\mbox{e}^{\frac{\hbar(\omega-\omega_{0,b})}{k_{B}T_{+}}}-1}\ .
\label{eq:espectro_rad_Kerr-Newman-Kasuya_case2_x_2_2}
\end{equation}
This is exactly the resulting Hawking radiation spectrum for scalar particles being radiated from a Kerr-Newman-Kasuya black hole with a cosmic string passing through it, where the Boltzmann's and Planck's constants were reintroduced.

Therefore, we can see that the resulting Hawking radiation spectrum of scalar particles has a thermal character, analogous to the black body spectrum, where $k_{B}T_{+}=\hbar\kappa_{+}/2\pi$, with $k_{B}$ being the Boltzmann constant. It is worth noticing that the total energy of radiated scalar particles decreases due to the presence of the cosmic string, more precisely, the dragging angular velocity of the exterior horizon, $\Omega_{+,b}$, is amplified in comparison with the scenario without a cosmic string \cite{AnnPhys.350.14}.

By integrating the above spectra or distribution function over all $\omega$'s, we can obtain the Hawking flux for massive scalar particles. It is given by
\begin{eqnarray}
\mbox{Flux} & = & \frac{1}{2\pi}\int_{0}^{\infty}{\bar{N}_{\omega}\ \omega\ d\omega}\nonumber\\
& = & \frac{1}{2\pi}\int_{0}^{\infty}{\frac{\omega\ d\omega}{\mbox{e}^{\frac{2\pi}{\kappa_{+}}(\omega-\omega_{0,b})}-1}}\nonumber\\
& = & \frac{1}{2\pi}\left(\frac{\kappa_{+}}{2\pi}\right)^{2}\mbox{Li}_{2}(\mbox{e}^{\frac{2\pi}{\kappa_{+}}\omega_{0,b}})\nonumber\\
& = & \frac{1}{2\pi}\left(\frac{\kappa_{+}}{2\pi}\right)^{2}\biggl\{\frac{\pi^{2}}{6}+\log(\mbox{e}^{\frac{2\pi}{\kappa_{+}}\omega_{0,b}})-\log(\mbox{e}^{\frac{2\pi}{\kappa_{+}}\omega_{0,b}})\log[-\log(\mbox{e}^{\frac{2\pi}{\kappa_{+}}\omega_{0,b}})]\nonumber\\
& + & \sum_{j=2}^{\infty}{\frac{\log^{j}(\mbox{e}^{\frac{2\pi}{\kappa_{+}}\omega_{0,b}})\zeta(2-j)}{j!}}\biggl\}\ ,
\label{eq:flux_rad_Kerr-Newman-Kasuya_case2_x_2}
\end{eqnarray}
where $\zeta(z)$ is the Riemann zeta function, and $\mbox{Li}_{n}(z)$ is the polylogarithm function with $n$ running from 1 to $\infty$. This is the exact result for the energy flux in the general case. In particular, for the Schwarzschild black hole, $\omega_{0,b}=0$, so we get
\begin{equation}
\mbox{Flux}=\frac{1}{2\pi}\left(\frac{\kappa_{+}}{2\pi}\right)^{2}\frac{\pi^{2}}{6}=\frac{\kappa_{+}^{2}}{48\pi}\ ,
\label{eq:flux_rad_Schwarzschild_case2_x_2}
\end{equation}
which is result already formally obtained in the literature \cite{PhysRevD.60.104013,PhysLettB.675.243}.
%
%
\section{Conclusions}
We have presented exact and general solutions for both angular and radial parts of the covariant Klein-Gordon equation for a charged massive scalar field in the Kerr-Newman-Kasuya spacetime (dyonic black hole) with a cosmic string passing through it and situated along its axis of symmetry.

These solutions extend the ones obtained in \cite{ClassQuantumGrav.23.7063}, if we take $Q_{m}=0$, in the sense that now we have analytic solutions for all spacetime, which means, in the region between the event horizon and infinity, differently from the results obtained in \cite{ClassQuantumGrav.23.7063} which are valid only in asymptotic regions, namely, very close to the horizons and far from the black hole. The radial solution is given in terms of the confluent Heun functions, and is valid over the range $0 \leq x < \infty$. As to the solution of the angular part, it also is given in terms of the confluent Heun functions.

From these analytic solutions, corresponding to the radial part, we obtained the solutions for ingoing and outgoing waves near the exterior horizon in the spacetime under consideration, and used these results to discuss the Hawking radiation effect, by taking into account the properties of the confluent Heun functions. This approach has the advantage that it is not necessary to introduce any coordinate system, as for example, the particular ones used in \cite{KexueTongbao.29.1303,ChinPhysLett.22.2492,ChinPhysB.17.2321}. As the dragging angular velocity of the exterior horizon, $\Omega_{+,b}$, depends on the conicity, this quantity codifies the presence of the cosmic string, and in fact, is amplified by the presence of this topological defect.

We also obtain the general expression for the energy flux for charged massive scalar particles, which is given in terms of the polylogarithm function.

From the local point of view, the gravitational field associated with the Kerr-Newman-Kasuya black hole with a cosmic string remains axially symmetric. But globally this symmetry was broken, and this fact produces some modifications in the physical states of the particles. The wave function depends on the parameter $b$ that codifies the presence of the string, and as a consequence all other physical quantities are also influenced by the presence of the cosmic string.
%
%
\ack The authors would like to thank Conselho Nacional de Desenvolvimento Cient\'{i}fico e Tecnol\'{o}gico (CNPq) for partial financial support.
%
%

%
%
\end{document}